\documentclass{article}
\usepackage{spconf,amsmath,graphicx,hyperref}
\usepackage{inconsolata}
\usepackage{booktabs}
\usepackage{multirow}
\usepackage{siunitx}
\usepackage[utf8]{inputenc}   %
\usepackage[T1]{fontenc}      %
\usepackage{booktabs}         %
\usepackage{amsmath,amssymb}  %
\usepackage{enumitem}         %

\usepackage{cleveref}

\usepackage{subcaption}
\usepackage{pifont}
\newcommand{\cmark}{\ding{51}}%
\newcommand{\xmark}{\ding{55}}%
\usepackage{todonotes}
\usepackage[most]{tcolorbox} %
\usepackage{xcolor}
\usepackage{array}
\usepackage{colortbl}

\usepackage{framed}

\usepackage{pifont}
\renewcommand{\cmark}{\textcolor{green!60!black}{\ding{51}}}
\renewcommand{\xmark}{\textcolor{red!80!black}{\ding{55}}}

\usepackage{placeins}  %
\usepackage{float}     %

\title{MusiCRS: Benchmarking Audio-Centric Conversational Recommendation}

\name{%
  \begin{tabular}{c}
  Rohan Surana$^{\star}$ \quad Amit Namburi$^{\star}$ \quad
  Gagan Mundada$^{\star}$ \quad Abhay Lal$^{\star}$\\
  Zachary Novack \quad Julian McAuley \quad Junda Wu
  \end{tabular}
  \thanks{$^{\star}$\,Equal contribution.}
}

\address{University of California, San Diego, USA}
 
\begin{document}
\ninept
\maketitle

\begin{abstract}
Conversational recommendation has advanced rapidly with large language models (LLMs), yet music remains a uniquely challenging domain in which effective recommendations require reasoning over audio content beyond what text or metadata can capture.
We present \textbf{MusiCRS}, the first benchmark for audio-centric conversational recommendation that links \emph{authentic} user conversations from Reddit with corresponding tracks. MusiCRS includes 477 high-quality conversations spanning diverse genres (classical, hip-hop, electronic, metal, pop, indie, jazz), with 3,589 unique musical entities and audio grounding via YouTube links. 
MusiCRS supports evaluation under three input modality configurations: audio-only, query-only, and audio+query, allowing systematic comparison of audio-LLMs, retrieval models, and traditional approaches. 
Our experiments reveal that current systems struggle with cross-modal integration, with optimal performance frequently occurring in single-modality settings rather than multimodal configurations.
This highlights fundamental limitations in cross-modal knowledge integration, as models excel at dialogue semantics but struggle when grounding abstract musical concepts in audio. To facilitate progress, we release the MusiCRS dataset\footnote{\url{https://huggingface.co/datasets/rohan2810/MusiCRS}}, evaluation code\footnote{\url{https://github.com/rohan2810/musiCRS}}, and comprehensive baselines.

\end{abstract}


\section{Introduction}
\label{sec:intro}

\begin{figure}[t]
  \centering
  \includegraphics[width=\columnwidth]{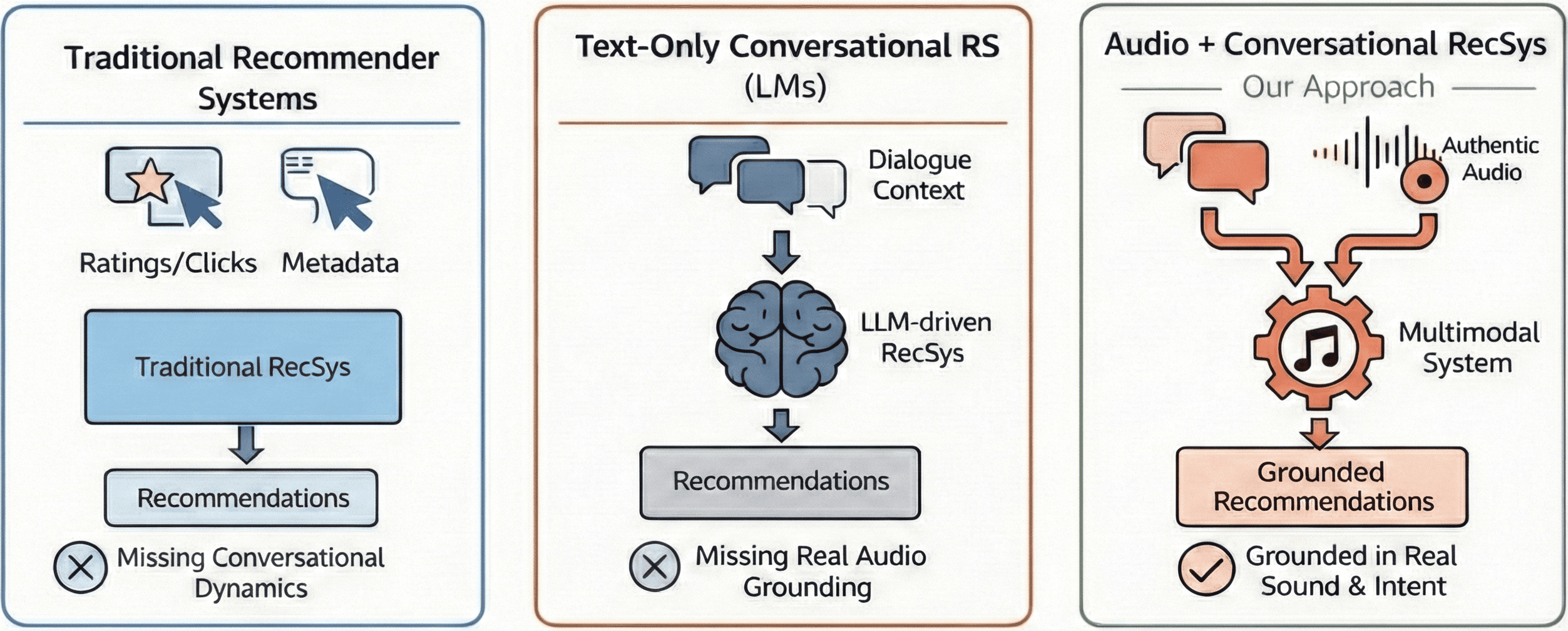}
  \vspace{0.5em}
  \scriptsize
  \resizebox{\columnwidth}{!}{%
  \begin{tabular}{l l c c c c c}
  \toprule
  \textbf{Dataset} & \textbf{Source Type} & \textbf{Real} & \textbf{Audio} & \textbf{Ground.} & \textbf{Rec.} & \textbf{Multi.} \\
  \midrule
  CPCD (Playlist) \cite{chaganty2023beyond} & Crowdsourced & \cmark & \xmark & \cmark & \cmark & \xmark \\
  TalkTheWalk \cite{leszczynski2023talk}    & Synthetic (LLM+lists) & \xmark & \xmark & \cmark & \cmark & \xmark \\
  LP-MusicDialog \cite{doh2024music}        & Synthetic (LLM-gen) & \xmark & \xmark & \cmark & \cmark & \xmark \\
  Audio Dialogues \cite{goel2024audio}      & Synthetic (captions) & \xmark & \cmark & \cmark & \xmark & \cmark \\
  LastFM \cite{Bertin-Mahieux2011}          & User logs & \xmark & \xmark & \xmark & \cmark & \xmark \\
  \midrule
  \textbf{MusiCRS} & In-the-wild (Reddit) & \cmark & \cmark & \cmark & \cmark & \cmark \\
  \bottomrule
  \end{tabular}}
  \caption{Limitations of existing conversational recommendation approaches (top) and comparison of music recommendation datasets (bottom). MusiCRS is the only benchmark combining authentic conversations, audio grounding, ground truth annotations, recommendation evaluation, and multimodal capabilities.}
  \label{fig:overview-datasets}
\end{figure}

Traditional music recommender systems rely on collaborative filtering and metadata but ignore the conversational dynamics through which users naturally express preferences. Conversational recommender systems powered by large language models (LLMs) address this limitation, enabling multi-turn interactions that track user intent and produce contextually relevant recommendations. While substantial progress has been made in domains like movies and e-commerce~\cite{hou2024bridging,li2018towards,jannach2021survey}, music recommendation presents unique challenges. Listener preferences depend not only on dialogue context but also on musical characteristics such as rhythm, timbre, instrumentation, and production style. These subtle musical dimensions are difficult to capture through text or metadata alone~\cite{mundada2025wildscore}, making music an ideal testbed for multimodal conversational recommendation. Current benchmarks predominantly focus on movies or general domains~\cite{zhang2018personalizing}, while music datasets emphasize metadata, tags, or listening logs~\cite{Bertin-Mahieux2011} rather than conversations referencing actual recordings. This gap prevents the evaluation of whether systems can bridge natural conversational cues with the musical characteristics driving human preference, limiting progress in multimodal music understanding.

\begin{figure}[t]
  \centering
  \includegraphics[width=\columnwidth]{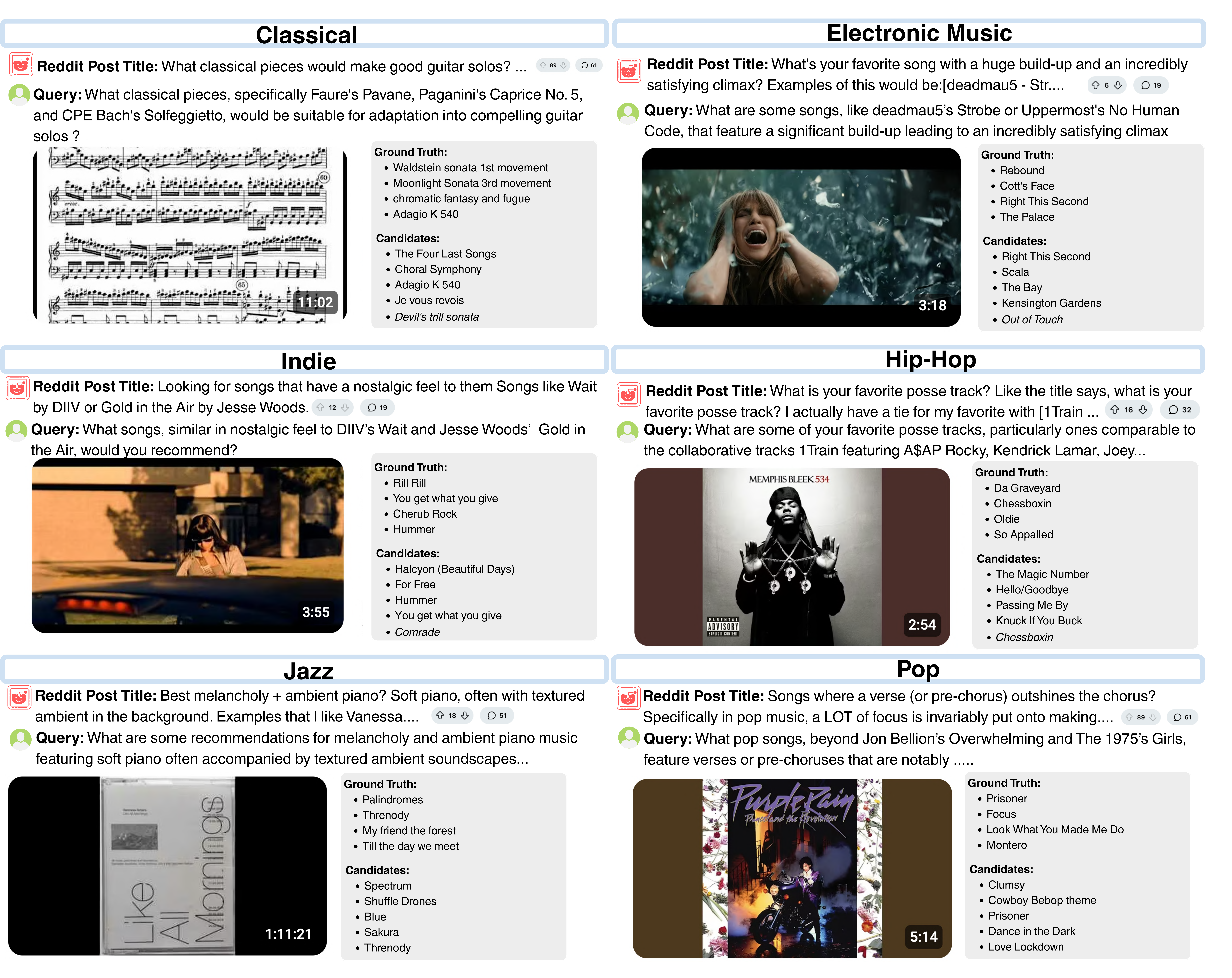}
  \caption{Representative examples from MusiCRS showing Reddit conversations, derived queries, audio, and candidates across genres.}
  \label{fig:exam}
\end{figure}

To address this gap, we introduce \textbf{MusiCRS}, the first benchmark for audio-centric conversational recommendation that links conversations with corresponding audio tracks. Built from large-scale Reddit discussions where users naturally exchange music suggestions, the dataset grounds each query in a referenced track, reflecting real-world music discovery. As illustrated in Fig.~\ref{fig:overview-datasets}, existing approaches face complementary limitations: traditional systems ignore conversational dynamics, while text-only conversational systems miss real audio grounding. MusiCRS addresses both by integrating dialogue context with authentic audio within a single benchmark. MusiCRS contains 477 high-quality Reddit discussions spanning seven genres, with each conversation grounded in audio through validated YouTube links. This design enables rigorous evaluation of cross-modal understanding in conversational music recommendation.

MusiCRS enables evaluation across three input modality configurations that systematically assess the contribution of different information sources. The \emph{audio-only} configuration tests models using only the referenced audio tracks, isolating the impact of musical content. The \emph{query-only} configuration evaluates models relying solely on conversational text, aligned with traditional text-based recommendation systems. The \emph{audio+query} configuration combines both modalities, testing recent advances in audio-LLMs~\cite{Qwen2-Audio,tang2023salmonn,goel2025audio} and multimodal retrieval models~\cite{elizalde2023clap,wu2025collap}. This design enables systematic analysis of how different information sources contribute to music recommendation quality across generative, retrieval-based, and traditional approaches. Our contributions are threefold:

\begin{itemize}
\item We introduce \textbf{MusiCRS}, the first benchmark for audio-centric conversational music recommendation, built from 477 Reddit discussions across seven genres with comprehensive audio grounding via validated YouTube links.
\item We establish rigorous evaluation protocols across three input modalities and provide comprehensive benchmarking of audio-LLMs, retrieval models, and traditional approaches, revealing substantial performance gaps in current systems.
\item We uncover a fundamental limitation in current multimodal models: optimal performance frequently occurs in single-modality settings, showing that models fail to ground abstract musical concepts in audio content.
\end{itemize}

\section{MusiCRS Dataset}
\label{sec:dataset}

\subsection{Dataset Construction}

\textbf{Large-Scale Reddit Mining and Filtering.} We collected 2.7 million Reddit submissions and 28.5 million comments from high-activity, music-oriented subreddits, including \textit{r/classicalmusic}, \textit{r/hiphopheads}, \textit{r/electronicmusic}, \textit{r/metal}, \textit{r/popheads}, \textit{r/indieheads}, and \textit{r/jazz}, following prior large-scale Reddit-based efforts in conversational AI~\cite{baumgartner2020pushshift,yang2018response}. We applied a three-stage filtering pipeline to ensure both musical grounding and conversational quality. First, only posts with valid YouTube links were retained, yielding $46,218$ threads linked to external music resources. Second, threads with fewer than three first-level replies or comments under five characters were removed, leaving $10,167$ multi-turn discussions with substantive context~\cite{zhang2018personalizing}. Finally, we discarded low-interaction threads during human annotation, ensuring the final dataset contained coherent and contextually rich exchanges.

\textbf{Music Entity Extraction and Query Generation.} We used LLMs to transform Reddit discussions into structured inputs for downstream tasks. Entity extraction was performed with \texttt{Qwen2.5-7B}~\cite{qwen2.5}, identifying and categorizing songs, artists, and albums from post titles, bodies, and top-level comments. In parallel, \texttt{gemma-3-12b-it}~\cite{team2025gemma} generated concise, context-preserving queries from each post, capturing user intent to support recommendation and retrieval.

\textbf{Query Normalization and Enhanced Candidate Selection.} For evaluation, we constructed candidate pools that include the ground truth entities while remaining realistic. For each conversation, we take the top $10$ ground-truth entities (ranked by comment upvotes) and add about $90$ additional entities randomly sampled from the same subreddit. We then shuffle the pool to remove positional bias. This design provides broad evaluation coverage while preserving topical coherence within each music community.

\textbf{Corpus Refinement and Audio Grounding.}  We refine the dataset by retaining conversations with $2$–$30$ musical entities, focusing on substantive discussions while eliminating noise. We cap each comment at $5$ entities to prevent verbosity and topic drift~\cite{xu2024cleaner}. Ground truth entities are extracted from comments and ranked by upvote score. We use upvotes as a popularity signal while maintaining diverse subreddit representation, balancing mainstream and niche musical content~\cite{salganik2023fairness}. YouTube links were validated and segmented into audio clips, grounding each conversation in authentic music content for evaluation. Fig.~\ref{fig:exam} illustrates genre-specific conversations, including the derived query, audio clip, and candidate pools.

\textbf{Human Validation.} To ensure conversational quality and audio grounding, four graduate student annotators manually validated $1,623$ conversations. Each was categorized as \textit{music}, \textit{pop culture}, or \textit{other}, including YouTube link verification and entity relevance review. During this process, $41.3\%$ of URLs were discarded as invalid, while $10.5\%$ were corrected or added through manual verification. Entity refinement removed $20.0\%$ of irrelevant entities and recovered $0.3\%$ of missing entities, improving overall precision. Additionally, $31.1\%$ of conversations were reassigned to the correct categories, ensuring proper alignment with the music-focused benchmark. Spam and off-topic content were removed throughout this validation process. For the final MusiCRS benchmark, we retained only the conversations classified as \textit{music}, resulting in $477$ high-quality music-focused conversational threads.

\begin{figure}[t]
  \centering
  \includegraphics[width=\columnwidth]{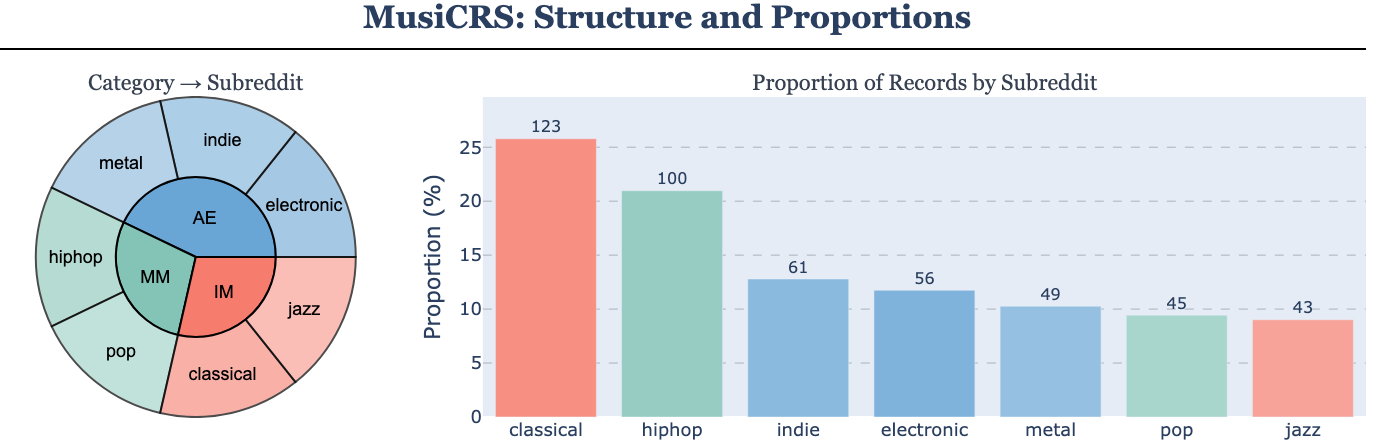}
  \caption{Dataset genre and entity coverage (left) and dialogue structure statistics (right). MusiCRS encompasses diverse genres, artists, and song distributions, and a wide range of conversation styles. \textbf{Macro-genre abbreviations:} MM (Modern \& Mainstream: \textit{hip-hop}, \textit{pop}); IM (Instrumental / Art Music: \textit{classical}, \textit{jazz}); AE (Alternative / Experimental: \textit{indie}, \textit{metal}, \textit{electronic}).}
  \label{fig:sidebyside_fullwidth}
\end{figure}

\subsection{Dataset Statistics}

The curated MusiCRS benchmark contains $477$ conversational threads spanning $3,589$ musical entities, with an average of $8$ ground truth entities per thread. Each thread includes $100$ evaluation candidates, constructed by combining the ground-truth items with subreddit-specific distractors to form realistic recommendation lists. Fig.~\ref{fig:sidebyside_fullwidth} summarizes genre balance and subreddit proportions: Classical (25.8\%), Hip-Hop (21.0\%), Indie (12.8\%), Electronic (11.7\%), Metal (10.3\%), Pop (9.4\%), and Jazz (9.0\%). This distribution ensures comprehensive coverage across mainstream and niche musical content, capturing variation in conversational styles, knowledge depth, and recommendation contexts, establishing MusiCRS as a robust benchmark for conversational, multimodal, and content-based music recommendation.

\begin{tcolorbox}[
  title=Case Study (RQ1), colback=gray!5,  colframe=black!70, fonttitle=\bfseries\sffamily,
  boxrule=0.5pt, arc=2pt, left=4pt, right=4pt, top=2pt, bottom=2pt
]\label{box:rq1}
\footnotesize
\sffamily

\textbf{Q:} What songs evoke overwhelming sorrow and emptiness, similar to Mozart's Lacrimosa or the opening of Elfen Lied?

\vspace{2pt}
\textbf{Ground Truth:}
Nymphs in the woods, Lamentations of Jeremiah, Ne irascaris \& Civitas, The One, Elfen Lied, Lost in Paradise, Speak to Me, Adagio, Miserere, 2nd movement Andante
\vspace{2pt} \\
\textbf{Query (Top-5):}\\
Wake Up, Numb, My Immortal, One More Light, Fix You\\[2pt]
\textbf{Audio (Top-5):}\\
Ave Maria, Clair de Lune, Moonlight Sonata, Canon in D, Adagio for Strings\\[2pt]
\textbf{Combined (Top-5):}\\
\textbf{Lamentations of Jeremiah}, Daphnis et Chlo\'{e}, \textbf{Elfen Lied}, Requiem, \textbf{Miserere}

\end{tcolorbox}

\section{Experiments}
To evaluate the effectiveness of audio-centric conversational recommendation systems on the \textbf{MusiCRS} dataset, we conduct comprehensive experiments across multiple model paradigms and input modalities. Our analysis focuses on three key dimensions: (1) comparing generative, retrieval-based, and traditional recommendation methods; (2) evaluating the impact of different input modalities (audio, text query, and multimodal combinations); and (3) assessing model performance across diverse musical genres and subreddits.

\begin{tcolorbox}[
  title=Case Study (RQ2), colback=gray!5, colframe=black!70, fonttitle=\bfseries\sffamily,
  boxrule=0.5pt, arc=2pt, left=4pt, right=4pt, top=2pt, bottom=2pt
] \label{box:rq2}
\footnotesize
\sffamily
\textbf{Q:} Following a recommendation from a Benjamin Zander TED Talk, what classical works would you suggest for a newcomer seeking emotionally evocative pieces?

\vspace{2pt}
\textbf{Ground Truth:}
Adagio for Strings, Symphony No.6, 7th Symphony (Second Movement), Moonlight Sonata, Op.18, Six Etudes-tableaux, Adagio-Lacrimosa, Elgar Cello Concerto, Nocturne Op. 55 No.1 in F minor, Ballade in G Minor

\vspace{2pt}
\textbf{Traditional (Popularity) (Top-5):}\\
Fantasy in F minor, Rosamunde Quartetsatz, \textbf{Nocturne Op. 55 No. 1 in F minor}, Late-Romantic, Lux Aurumque\\[2pt]
\textbf{Retrieval (CLAP) (Top-5):}\\
Waldstein Sonata, Shostakovich Concerto, \textbf{Six Etudes-tableaux}, \textbf{7th Symphony (Second Movement)},Piano Concerto No.2 Mvt.3\\[2pt]
\textbf{Generative (Qwen2.5-Omni) (Top-5):}\\
\textbf{Adagio for Strings}, Rach 1 Ossia, \textbf{Ballade in G Minor}, \textbf{7th Symphony (Second Movement)}, La Mer
\end{tcolorbox}

\subsection{Experimental Setup}

We frame experiments as ranking tasks across three modalities: \textit{audio-only}, \textit{query-only}, and \textit{audio+query}. Each query uses a $300$-second audio budget across up to $10$ clips ($30$ seconds for SALMONN). Audio is resampled to model-appropriate rates ($16$kHz for SALMONN, $48$kHz for CLAP). We employ Recall, nDCG, and MRR as primary metrics, aggregated across the dataset and per-subreddit. Generative models use default generative settings with increased \emph{max\_new\_tokens} (typically $512$ tokens) to accommodate full candidate rankings. Retrieval models compute cosine similarity between query embeddings and candidate song embeddings, with late fusion for multimodal scenarios. We evaluate seven generative models (Qwen2-Audio-7B~\cite{Qwen2-Audio}, Qwen2.5-Omni-7B~\cite{xu2025qwen2}, SALMONN-7B/13B~\cite{tang2023salmonn}, FUTGA~\cite{wu2024futga}, Audio Flamingo 3~\cite{goel2025audio}, Phi-4-Multimodal~\cite{abouelenin2025phi}), two retrieval systems (CLAP~\cite{wu2023large}, CoLLAP~\cite{wu2025collap}), and two traditional baselines (subreddit-specific popularity, neighborhood-based recommendation).

\begin{table*}[htb]
\footnotesize
\centering
\begin{tabular}{@{}ll*{8}{c}@{}}
\toprule
\textbf{Model} & \textbf{Modality} & {\textbf{Classical}} & {\textbf{Hip-Hop}} & {\textbf{Pop}} & {\textbf{Electronic}} & {\textbf{Metal}} & {\textbf{Indie}} & {\textbf{Jazz}} & {\textbf{Overall}} \\
\midrule
\multicolumn{10}{c}{\cellcolor{blue!12}\textbf{Generative Models}} \\
\midrule
\multirow{3}{*}{\textbf{Qwen2-Audio-7B}} 
  & Audio  & 15.83/12.10 & 14.20/11.40 & 17.47/13.20 & 17.90/11.75 & 18.01/11.02 & 16.74/12.48 & 19.19/10.98 & 16.53/11.86 \\
  & Query  & 16.15/10.97 & 17.16/13.01 & 18.02/13.33 & 10.70/07.98  & 13.73/10.05 & 17.18/13.86 & 16.00/10.72 & 15.77/11.52 \\
  & Combined & 16.38/12.71 & 18.39/14.03 & 15.32/12.29 & 14.58/10.23 & 17.40/13.26 & 15.61/12.63 & 21.46/13.76 & \textbf{16.95/12.80} \\
\cmidrule(lr){1-10}

\multirow{3}{*}{\textbf{Qwen2.5-Omni-7B}} 
  & Audio  & 14.80/10.79 & 19.24/13.28 & 16.47/12.07 & 19.79/15.22 & 22.47/13.57 & 21.08/16.62 & \underline{28.09/16.36} & 19.26/13.48 \\
  & Query & 14.07/12.04 & 18.91/14.10 & \underline{23.38/17.59} & 18.00/13.50 & 16.45/12.62 & 20.39/14.49 & 22.48/16.73 & 18.24/13.96 \\
  & Combined & \underline{26.53/19.33} & 17.74/13.55 & 18.92/13.14 & \underline{25.29/19.03} & {23.61}/16.91 & 17.30/14.23 & 22.00/15.02 & \textbf{21.93/16.21} \\
\cmidrule(lr){1-10}

\multirow{3}{*}{\textbf{SALMONN-7B}} 
  & Audio  & 17.88/12.89 & \underline{25.12/16.85} & 15.41/10.66 & 16.79/12.16 & 21.41/15.08 & 22.24/17.07 & 20.76/14.35 & \textbf{20.22}/14.31 \\
  & Query & 19.17/13.01 & 19.33/14.05 & 14.07/9.68 & 23.55/13.66 & 17.83/12.60 & 19.22/13.55 & 13.49/8.63 & 18.60/12.62 \\
  & Combined & 21.70/15.06 & 19.91/13.15 & 15.74/11.71 & 18.31/12.97 & 16.03/10.60 & 18.99/15.75 & 23.35/15.06 & 19.58/13.73 \\
\cmidrule(lr){1-10}
\multirow{3}{*}{\textbf{SALMONN-13B}} 
  & Audio  & 17.12/12.08 & 22.31/15.14 & 16.39/11.13 & 18.12/12.31 & 19.15/14.57 & 19.94/14.98 & 25.18/16.26 & \textbf{19.55/13.66} \\
  & Query & 17.96/12.62 & 19.78/13.90 & 15.38/10.63 & 22.32/13.08 & 19.81/14.34 & 21.91/15.99 & 17.80/10.95 & 19.29/13.21 \\
  & Combined & 18.48/12.86 & 19.23/12.57 & 16.41/11.29 & 17.48/12.50 & 16.98/10.64 & 18.70/15.57 & 25.47/17.39 & 18.83/13.13 \\
  
\cmidrule(lr){1-10}
  
\multirow{3}{*}{\textbf{FUTGA}} 
  & Audio  & 18.89/11.72 & 17.84/12.25 & 19.25/13.69 & 17.60/12.72 & 19.79/12.53 & 22.60/14.98 & 20.32/14.07 & \textbf{19.25/12.84} \\
  & Query & 19.11/12.08 & 18.42/11.85 & 22.65/14.82 & 14.97/09.79 & 16.96/10.82 & 20.77/14.23 & 20.73/10.40 & 18.95/12.02 \\
  & Combined & 19.58/11.89 & 17.56/11.61 & 18.89/12.87 & 19.25/12.29 & 15.36/10.51 & 20.48/13.96 & 18.88/11.24 & 18.67/12.04 \\  
\cmidrule(lr){1-10}

\multirow{3}{*}{\textbf{Audio Flamingo 3}} 
  & Audio  & 17.30/12.38 & 14.20/10.78 & 13.61/10.33 & 16.02/11.79 & 22.41/12.35 & 21.62/15.54 & 19.48/12.88 & 17.42/12.23 \\
  & Query & 17.02/13.52 & 17.77/12.49 & 17.61/12.62 & 17.59/12.93 & 19.33/14.71 & 20.77/14.38 & 20.72/12.91 & \textbf{18.35/13.33} \\
  & Combined & 16.38/12.93 & 17.56/12.12 & 17.01/12.69 & 14.82/10.73 & 16.70/10.33 & 20.97/15.00 & 25.52/14.93 & 17.95/12.66 \\
\cmidrule(lr){1-10}

\multirow{3}{*}{\textbf{Phi-4-Multimodal}} 
  & Audio  & 21.86/15.08 & 20.72/14.57 & 14.56/10.01 & 18.17/13.53 & 18.75/11.38 & \underline{22.95/16.53} & 18.72/10.68 & \textbf{20.04/13.72} \\
  & Query & 24.38/18.57 & 17.01/13.37 & 16.69/11.45 & 12.82/08.58 & 15.82/10.70 & 29.85/17.40 & 17.31/10.52 & 19.93/13.95 \\
  & Combined & 19.94/12.96 & 16.91/11.63 & 17.91/12.59 & 18.23/12.59 & 15.80/09.79 & 20.68/14.80 & 22.31/15.72 & 18.79/12.76 \\

\midrule
\multicolumn{10}{c}{\cellcolor{orange!12}\textbf{Retrieval Models}} \\
\midrule

\multirow{3}{*}{\textbf{CLAP}} 
  & Audio  & 22.38/15.68 & 21.22/15.14 & 18.86/14.93 & 15.12/12.11 & \underline{26.42/16.25} & 20.56/15.41 & 22.51/13.40 & 21.15/14.90 \\
  & Query & 23.56/16.87 & 23.45/16.67 & 18.50/13.78 & 18.23/12.82 & 23.74/15.78 & {23.43}/17.09 & {26.56}/15.98 & \underline{\textbf{22.71/15.90}} \\
  & Combined & 23.79/16.96 & 22.10/15.85 & 18.52/15.24 & 17.26/12.19 & 25.00/16.43 & 22.14/17.18 & 27.61/15.23 & 22.43/15.82 \\
\cmidrule(lr){1-10}
\multirow{3}{*}{\textbf{CoLLAP}} 
  & Audio  & 17.23/12.22 & 19.90/14.49 & 19.75/13.52 & 19.04/12.75 & 20.24/14.11 & 18.17/14.13 & {27.75}/15.52 & 19.62/13.62 \\
  & Query & 19.26/12.70 & 20.30/14.75 & 20.38/15.96 & 22.11/13.87 & 20.19/12.07 & 21.65/16.44 & 25.14/14.35 & \textbf{20.85/14.14} \\
  & Combined & 18.68/12.86 & 21.09/15.08 & 21.88/15.98 & 21.17/14.63 & 22.35/13.04 & 18.42/12.66 & 23.08/14.02 & 20.52/13.92 \\  
\midrule
\multicolumn{10}{c}{\cellcolor{gray!12}\textbf{Traditional Methods}} \\
\midrule
\textbf{Popularity} & Query & 16.41/11.21 & 15.15/10.18 & 16.52/12.30 & 16.73/10.10 & 14.92/9.62 & 15.05/11.67 & 23.59/13.70 & 16.51/11.09 \\
\cmidrule(lr){1-10}
\textbf{Neighborhood} & Query & 20.12/13.53 & 17.47/10.92 & 15.85/10.35 & 13.76/7.92 & 11.91/6.67 & 8.96/5.59 & 4.30/2.38 & 14.72/9.30 \\

\bottomrule
\end{tabular}
\caption{Performance comparison across music genres and input modalities. Values represent Recall@20/nDCG@20. \textbf{Bold} values indicate the best-performing modality configuration for each model, and \underline{underlined} values show the top-performing model for each genre, both determined by Recall@20.}
\label{tab:performance_results}
\end{table*}

\subsection{Research Questions and Analysis}

Our comprehensive evaluation addresses three key research questions that illuminate fundamental challenges in multimodal music recommendation (Table~\ref{tab:performance_results}):

\textbf{RQ1: How do different input modalities impact music recommendation performance?} 

Our analysis reveals that multimodal approaches do not consistently outperform single-modality configurations. CLAP achieves optimal performance with query-only input ($22.71$\% Recall@20) compared to audio-only ($21.15$\%), while Qwen2.5-Omni demonstrates superior performance with audio+query combination ($21.93$\% vs. $19.26$\% audio-only and $18.24$\% query-only). However, only 2 of 9 multimodal-capable models achieve their best performance in multimodal settings. 
This finding aligns with recent work~\cite{weck2024muchomusic,zang2025listening} demonstrating that current multimodal models exhibit architectural limitations in cross-modal integration, suggesting that existing fusion mechanisms inadequately exploit complementary information across modalities. A representative example is provided in Box~\ref{box:rq1}.

\tcbset{
  casestudy/.style={
    enhanced,
    sharp corners,
    colback=white,
    colframe=black,
    coltitle=white,
    colbacktitle=black,
    fonttitle=\bfseries,
    boxrule=0.5pt,
    title filled,
    top=2pt, bottom=2pt, left=4pt, right=4pt
  }
}

\begin{figure}[htbp]
  \centering
  \includegraphics[width=\columnwidth]{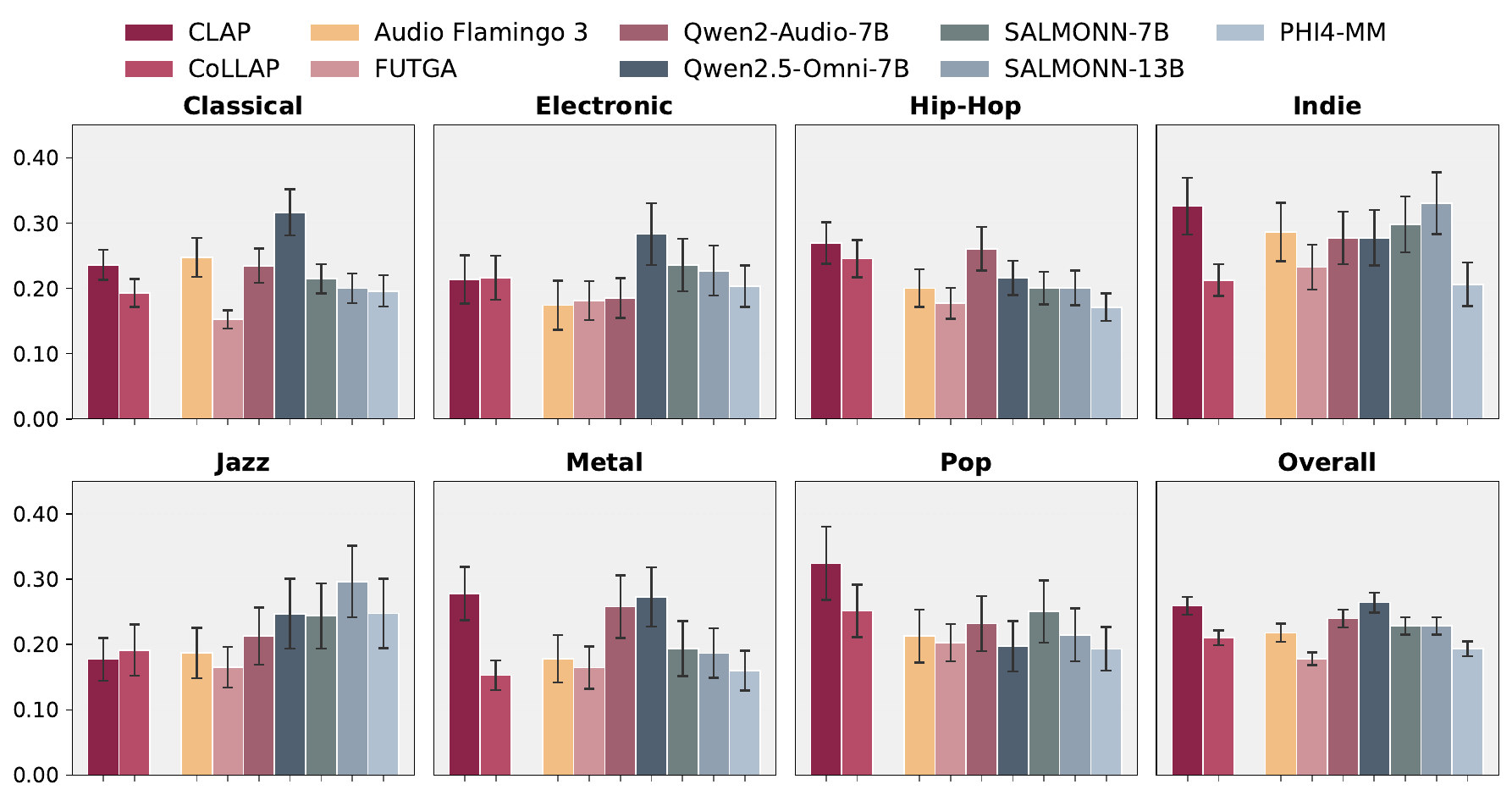}
  \caption{Mean Reciprocal Rank (MRR) comparison across genres and model types.}
  \label{fig:ana}
\end{figure}

\textbf{RQ2: How do different model paradigms compare in music recommendation?}

On average, retrieval-based approaches demonstrate superior performance 
compared to generative models across all metrics (Recall@20: $21.21$\% vs. 
$18.77$\%; nDCG@20: $14.72$\% vs. $13.14$\%).
CLAP achieves the highest overall performance ($22.71$\% Recall@20), while the best-performing generative model (Qwen2.5-Omni) reaches a competitive $21.93$\%, surpassing CoLLAP ($20.85$\%). Traditional methods exhibit substantially lower performance, with popularity-based recommendation achieving $16.51$\% and neighbourhood-based recommendation $14.72$\%, highlighting their inadequacy for audio-centric recommendation tasks that require understanding musical content beyond popularity signals. Within generative models, performance varies from $15.77$\% to $21.93$\% across modality configurations, whereas retrieval models range from $19.62$\% to $22.71$\%. These results indicate that learned embedding representations provide more robust recommendation signals than generative reasoning for music recommendation tasks; see Case Study in Box~\ref{box:rq2}.

\textbf{RQ3: What genre-specific patterns emerge?}

Performance analysis (Fig.~\ref{fig:ana}) reveals substantial genre-specific variation reflecting both domain-specific challenges and model capabilities. Jazz consistently achieves strong performance across most models ($28.09$\% Recall@20 for Qwen2.5-Omni audio-only; CLAP: $27.61$\% combined, CoLLAP: $27.75$\%), while Pop exhibits the lowest peak performance across genres ($23.38$\%). Metal shows notable model-dependent variation, with CLAP achieving $26.42$\% in audio-only configuration while other models show more moderate results (e.g., Qwen2.5-Omni combined: $23.61$\%). Classical demonstrates a distinctive pattern where multimodal approaches substantially outperform single-modality inputs (Qwen2.5-Omni: $26.53$\% combined vs.\ $14.80$\% audio-only and $14.07$\% query-only), suggesting that cross-modal integration may be particularly beneficial for this genre. These patterns indicate that genre-specific audio characteristics and linguistic conventions differentially impact model performance, with Jazz's distinctive musical features being more readily captured by current architectures.

\section{Conclusion}
We introduced \textbf{MusiCRS}, the first benchmark for audio-centric conversational recommendation grounded in both conversational context and corresponding audio. Extensive evaluations across recommendation paradigms and genres reveal key limitations of current multimodal and conversational models, particularly in nuanced audio-contextual reasoning. We release MusiCRS to support future advances in multimodal music recommendation.

\section{Acknowledgment}
This work was partially supported by the U.S. National Science Foundation under Grant IIS-2432486.
\bibliographystyle{IEEEbib}

\begin{thebibliography}{10}

\bibitem{chaganty2023beyond}
Arun~Tejasvi Chaganty, Megan Leszczynski, Shu Zhang, Ravi Ganti, Krisztian Balog, and Filip Radlinski,
\newblock ``Beyond single items: Exploring user preferences in item sets with the conversational playlist curation dataset,''
\newblock in {\em Proceedings of the 46th International ACM SIGIR Conference on Research and Development in Information Retrieval}, 2023, pp. 2754--2764.

\bibitem{leszczynski2023talk}
Megan Leszczynski, Shu Zhang, Ravi Ganti, Krisztian Balog, Filip Radlinski, Fernando Pereira, and Arun~Tejasvi Chaganty,
\newblock ``Talk the walk: Synthetic data generation for conversational music recommendation,''
\newblock {\em arXiv preprint arXiv:2301.11489}, 2023.

\bibitem{doh2024music}
SeungHeon Doh, Keunwoo Choi, Daeyong Kwon, Taesu Kim, and Juhan Nam,
\newblock ``Music discovery dialogue generation using human intent analysis and large language models,''
\newblock {\em arXiv preprint arXiv:2411.07439}, 2024.

\bibitem{goel2024audio}
Arushi Goel, Zhifeng Kong, Rafael Valle, and Bryan Catanzaro,
\newblock ``Audio dialogues: Dialogues dataset for audio and music understanding,''
\newblock {\em arXiv preprint arXiv:2404.07616}, 2024.

\bibitem{Bertin-Mahieux2011}
Thierry Bertin-Mahieux, Daniel~P.W. Ellis, Brian Whitman, and Paul Lamere,
\newblock ``The million song dataset,''
\newblock in {\em {Proceedings of the 12th International Conference on Music Information Retrieval ({ISMIR} 2011)}}, 2011.

\bibitem{hou2024bridging}
Yupeng Hou, Jiacheng Li, Zhankui He, An~Yan, Xiusi Chen, and Julian McAuley,
\newblock ``Bridging language and items for retrieval and recommendation,''
\newblock {\em arXiv preprint arXiv:2403.03952}, 2024.

\bibitem{li2018towards}
Raymond Li, Samira Ebrahimi~Kahou, Hannes Schulz, Vincent Michalski, Laurent Charlin, and Chris Pal,
\newblock ``Towards deep conversational recommendations,''
\newblock {\em Advances in neural information processing systems}, vol. 31, 2018.

\bibitem{jannach2021survey}
Dietmar Jannach, Ahtsham Manzoor, Wanling Cai, and Li~Chen,
\newblock ``A survey on conversational recommender systems,''
\newblock {\em ACM Computing Surveys (CSUR)}, vol. 54, no. 5, pp. 1--36, 2021.

\bibitem{mundada2025wildscore}
Gagan Mundada, Yash Vishe, Amit Namburi, Xin Xu, Zachary Novack, Julian McAuley, and Junda Wu,
\newblock ``Wildscore: Benchmarking mllms in-the-wild symbolic music reasoning,''
\newblock in {\em Proceedings of the 2025 Conference on Empirical Methods in Natural Language Processing}, 2025, pp. 16858--16874.

\bibitem{zhang2018personalizing}
Saizheng Zhang, Emily Dinan, Jack Urbanek, Arthur Szlam, Douwe Kiela, and Jason Weston,
\newblock ``Personalizing dialogue agents: I have a dog, do you have pets too?,''
\newblock {\em arXiv preprint arXiv:1801.07243}, 2018.

\bibitem{Qwen2-Audio}
Yunfei Chu, Jin Xu, Qian Yang, Haojie Wei, Xipin Wei, Zhifang Guo, Yichong Leng, Yuanjun Lv, Jinzheng He, Junyang Lin, Chang Zhou, and Jingren Zhou,
\newblock ``Qwen2-audio technical report,''
\newblock {\em arXiv preprint arXiv:2407.10759}, 2024.

\bibitem{tang2023salmonn}
Changli Tang, Wenyi Yu, Guangzhi Sun, Xianzhao Chen, Tian Tan, Wei Li, Lu~Lu, Zejun MA, and Chao Zhang,
\newblock ``{SALMONN}: Towards generic hearing abilities for large language models,''
\newblock in {\em The Twelfth International Conference on Learning Representations}, 2024.

\bibitem{goel2025audio}
Arushi Goel, Sreyan Ghosh, Jaehyeon Kim, Sonal Kumar, Zhifeng Kong, Sang-gil Lee, Chao-Han~Huck Yang, Ramani Duraiswami, Dinesh Manocha, Rafael Valle, and Bryan Catanzaro,
\newblock ``Audio flamingo 3: Advancing audio intelligence with fully open large audio language models,''
\newblock {\em arXiv preprint arXiv:2507.08128}, 2025.

\bibitem{elizalde2023clap}
Benjamin Elizalde, Soham Deshmukh, Mahmoud Al~Ismail, and Huaming Wang,
\newblock ``Clap learning audio concepts from natural language supervision,''
\newblock in {\em ICASSP 2023-2023 IEEE International Conference on Acoustics, Speech and Signal Processing (ICASSP)}. IEEE, 2023, pp. 1--5.

\bibitem{wu2025collap}
Junda Wu, Warren Li, Zachary Novack, Amit Namburi, Carol Chen, and Julian McAuley,
\newblock ``Collap: Contrastive long-form language-audio pretraining with musical temporal structure augmentation,''
\newblock in {\em ICASSP 2025-2025 IEEE International Conference on Acoustics, Speech and Signal Processing (ICASSP)}. IEEE, 2025, pp. 1--5.

\bibitem{baumgartner2020pushshift}
Jason Baumgartner, Savvas Zannettou, Brian Keegan, Megan Squire, and Jeremy Blackburn,
\newblock ``The pushshift reddit dataset,''
\newblock in {\em Proceedings of the International AAAI Conference on Web and Social Media (ICWSM)}, 2020, vol.~14, pp. 830--839.

\bibitem{yang2018response}
Liu Yang, Minghui Qiu, Chen Qu, Jiafeng Guo, Yongfeng Zhang, W~Bruce Croft, Jun Huang, and Haiqing Chen,
\newblock ``Response ranking with deep matching networks and external knowledge in information-seeking conversation systems,''
\newblock in {\em The 41st international acm sigir conference on research \& development in information retrieval}, 2018, pp. 245--254.

\bibitem{qwen2.5}
Qwen Team,
\newblock ``Qwen2.5: A party of foundation models,'' September 2024.

\bibitem{team2025gemma}
Gemma Team, Aishwarya Kamath, Johan Ferret, Shreya Pathak, Nino Vieillard, Ramona Merhej, Sarah Perrin, Tatiana Matejovicova, Alexandre Ram{\'e}, Morgane Rivi{\`e}re, et~al.,
\newblock ``Gemma 3 technical report,''
\newblock {\em arXiv preprint arXiv:2503.19786}, 2025.

\bibitem{xu2024cleaner}
Zhipeng Xu, Zhenghao Liu, Yukun Yan, Zhiyuan Liu, Ge~Yu, and Chenyan Xiong,
\newblock ``Cleaner pretraining corpus curation with neural web scraping,''
\newblock {\em arXiv preprint arXiv:2402.14652}, 2024.

\bibitem{salganik2023fairness}
Rebecca Salganik, Fernando Diaz, and Golnoosh Farnadi,
\newblock ``Fairness through domain awareness: Mitigating popularity bias for music discovery,''
\newblock {\em arXiv preprint arXiv:2308.14601}, 2023.

\bibitem{xu2025qwen2}
Jin Xu, Zhifang Guo, Jinzheng He, Hangrui Hu, Ting He, Shuai Bai, Keqin Chen, Jialin Wang, Yang Fan, Kai Dang, et~al.,
\newblock ``Qwen2. 5-omni technical report,''
\newblock {\em arXiv preprint arXiv:2503.20215}, 2025.

\bibitem{wu2024futga}
Junda Wu, Zachary Novack, Amit Namburi, Jiaheng Dai, Hao-Wen Dong, Zhouhang Xie, Carol Chen, and Julian McAuley,
\newblock ``Futga: Towards fine-grained music understanding through temporally-enhanced generative augmentation,''
\newblock {\em arXiv preprint arXiv:2407.20445}, 2024.

\bibitem{abouelenin2025phi}
Abdelrahman Abouelenin, Atabak Ashfaq, Adam Atkinson, Hany Awadalla, Nguyen Bach, Jianmin Bao, Alon Benhaim, Martin Cai, Vishrav Chaudhary, Congcong Chen, et~al.,
\newblock ``Phi-4-mini technical report: Compact yet powerful multimodal language models via mixture-of-loras,''
\newblock {\em arXiv preprint arXiv:2503.01743}, 2025.

\bibitem{wu2023large}
Yusong Wu, Ke~Chen, Tianyu Zhang, Yuchen Hui, Taylor Berg-Kirkpatrick, and Shlomo Dubnov,
\newblock ``Large-scale contrastive language-audio pretraining with feature fusion and keyword-to-caption augmentation,''
\newblock in {\em ICASSP 2023-2023 IEEE International Conference on Acoustics, Speech and Signal Processing (ICASSP)}. IEEE, 2023, pp. 1--5.

\bibitem{weck2024muchomusic}
Benno Weck, Benjamin Elizalde, Zhiyao Wang, Bhiksha Raj, and Brian McFee,
\newblock ``Muchomusic: Evaluating music understanding in multimodal audio-language models,''
\newblock in {\em International Society for Music Information Retrieval Conference}, 2024.

\bibitem{zang2025listening}
Yannis Zang, Benjamin Elizalde, Zhiyao Wang, and Brian McFee,
\newblock ``Are you really listening? boosting perceptual awareness in music-qa benchmarks,''
\newblock in {\em International Society for Music Information Retrieval Conference}, 2025.

\end{thebibliography}

\end{document}